\documentclass{article}
\usepackage{hiph-preprint}
\usepackage{amsmath}
\usepackage{epsfig}
\volnumber{22} \issuenumber{1} \edyear{2005}                             %|
\frompage{000} \topage{000}                                              %|
\recrevdate{1 January 2005}                                              %|
\def\rightmark{Flow and jet quenching power}
\def\leftmark{T.~Renk}
 
\title{The influence of flow on the jet quenching power in heavy-ion collisions} 
\authors{ 
{Thorsten Renk$^{1,2}$  %
}\\[2.812mm]
{\normalsize
\hspace*{-8pt}$^1$ Department of Physics, PO Box 35
FIN-40014 University of Jyv\"{a}skyl\"{a}, Finland\\[0.2ex] \\
\hspace*{-8pt}$^2$Helsinki Institute of Physics, PO Box 64
FIN-00014 University of Helsinki, Finland\\
}}
\abstract{The flow pattern and evolution of the medium created in ultrarelativistic heavy ion collisions
can have significant influence on the energy loss of hard partons traversing the medium. We demonstrate
that within a range of assumptions for longitudinal and transverse flow which are all compatible
with the measured hadronic single particle distributions, the quenching power of the medium can vary
within a factor five. Thus, the choice of the medium evolution is one of the biggest uncertainties
in jet quenching calculations and needs to be addressed with some care.}

 
\makeindex
\begin{document}
 
\maketitle

\section{Introduction}\label{intro}
Energy loss of a high $p_T$ 'hard' parton travelling through low $p_T$ 'soft' matter has long been 
recognized as a promising tool to study the initial high-density phases of ultrarelativistic
hevay-ion collisions (URHIC) \cite{Jet1,Jet2,Jet3,Jet4,Jet5,Jet6}. In \cite{Urs1,Urs2}, it
has been suggested that a flow component transverse to the high $p_T$ parton trajectory
would lead to increased energy loss as compared to the one a static medium. 
It seems that this energy loss ends up exciting
Mach cone like hydrodynamical shock waves traversing the medium \cite{Mach}.

In \cite{Jet_Flow} we investigated this suggestion in a hydro-inspired evolution model for 
Au-Au collisions at RHIC \cite{RenkSpectraHBT} which successfully reproduces hadronic 1-particle spectra
and 2-particle correlation measurements.

We found that if one wants to be consistent with the measured value of the nuclear suppression
factor $R_{AA}$ \cite{STAR_RAA}, the transport coefficient $\hat{q}$ (the parameter locally characterizing
the opacity of the medium) has to be readjusted within a factor 5 for different assumptions about the
flow profile and development. In this paper we summarize the essential findings of \cite{Jet_Flow}.

\section{The formalism}

Key quantity for the calculation of jet energy loss is the local transport coefficient
$\hat{q}(\eta_s, r, \tau)$ which characterizes the squared average momentum transfer
from the medium to the hard parton per unit pathlength. Since we consider a time-dependent
inhomogeneous medium, this quantity depends on spacetime rapidity $\eta_s = \frac{1}{2}\ln \frac{t+z}{t-z}$,
radius $r$ and proper time $\tau = \sqrt{t^2-z^2}$ (we focus on central collisions and
assume azimuthal symmetry in the following).
The transport coefficient is related to the energy density of the medium as
$\hat{q} = c \epsilon^{3/4}$.

In order to find the probability for a hard parton $P(\Delta E)$ to lose the energy
$\Delta E$ while traversing the medium, we make use of a scaling law \cite{JetScaling} which allows
to relate the dynamical scenario a static equivalent one by calculating the following quantities
averaged over the jet trajectory $\xi(\tau):$

\begin{equation}
\omega_c({\bf r_0}, \phi) = \int_0^\infty d \xi \xi \hat{q}(\xi)
\quad \text{and} \quad
(\hat{q}L) ({\bf r_0}, \phi) = \int_0^\infty d \xi \hat{q}(\xi)
\end{equation}

as a function of the jet production vertex ${\bf r_0}$ and its angular orientation $\phi$.
In the presence of flow, we follow the prescription suggested in \cite{Urs2} and replace

\begin{equation}
\label{E-Urs}
\hat{q} = c \epsilon^{3/4}(p) \rightarrow c \epsilon (T^{n_\perp  n_\perp})
\quad \text{with} \quad
T^{n_\perp n_\perp} = p(\epsilon) + \left[ \epsilon + p(\epsilon)\right] \frac{\beta_\perp^2}{1-\beta_\perp^2}
\end{equation}

where $\beta_\perp$ is the spatial component of the flow field orthogonal to the parton trajectory.
Using the results of \cite{QuenchingWeights}, we obtain $P(\Delta E)$ from $\omega_c$ and $(\hat{q}L)$
as a function of jet production vertex and the angle $\phi$ from the distribution $\omega \frac{dI}{d\omega}$ 
of gluons emitted into the jet cone.
 We average over all possible angles and
production vertices, weighting the distribution with the nuclear overlap $T_{AA}({\bf b}) = \int dz \rho^2({\bf b},z)$
(for central collisions) with $\rho$ the nuclear density as a function of impact parameter {\bf b} and longitudinal
coordinate $z$. 

We calculate the the inclusive charged pion production in LO pQCD. Thus, schematically this amounts to folding the
average energy loss probability into the factorized expression for hadron production (explicit expressions
can be found in \cite{Kari1,Kari2}).
We use the CTEQ6 parton distribution functions \cite{CTEQ1,CTEQ2} for the pp reference, the
NPDF set \cite{NPDF} for production in nuclear collisions and the KKP fragmentation functions \cite{KKP}.

We obtain the nuclear modification factor (in the case of central collisions) as
\begin{equation}
R_{AA}(p_T,y) = \frac{d^2N^{AA}/dp_Tdy}{T_{AA}(0) d^2 \sigma^{NN}/dp_Tdy}.
\end{equation}

In order to demonstrate the influence of flow, we calculate two different scenarios: 
In a first run, we assume a longitudinal Bjorken expansion (as commonly done) and disregard the correction Eq.~\ref{E-Urs}. Then, we use the best fit to the hadronic freeze-out from \cite{RenkSpectraHBT}, assume
a transverse flow profile $v_T \sim r^2$ and a small initial $v_T$ and take into account Eq.~\ref{E-Urs}. 
We stress that
both scenarios describe transverse mass distribution of pions, kaons and protons as well as $dN/d\eta$ spectra,
however the Bjorken expansion does not describe the HBT correlations correctly. To get a measure
for the opacity, we then adjust the coefficient $c$ linking $\epsilon^{3/4}$ and $\hat{q}$ such that the high $p_T$ tail of $R_{AA}$
(where the formalism is applicable) is described well.

\section{Results and discussion}

\begin{figure}[!htb]
\epsfig{file=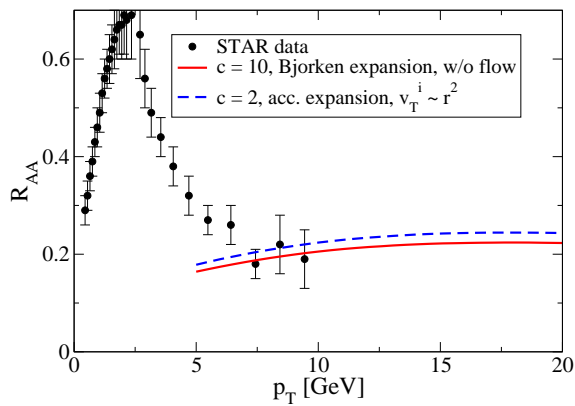, width=7.5cm}
\caption{\label{F-1}Calculated $R_{AA}$ in a Bjorken expansion and without the effect of transverse flow on
energy loss, requiring $c=10$ (solid red) and in the best fit evolution described in \cite{RenkSpectraHBT}, assuming
a quadratic flow profile, a small primordial flow velocity $v^i_T = 0.1$ and $\alpha_s = 0.45$.}
\end{figure}

The resulting description can be seen in Fig.~\ref{F-1}. We stress that $c$ varies bewteen 2 and 10, thus
flow potentially has a dramatic influence on the quenching power of the medium.

The physics behind the influence of longitudinal and transverse flow is quite different. At midrapidity,
jets are always co-moving with the surrounding matter, so Eq.\ref{E-Urs} is irrelevant for 
longitudinal flow. However, the density as a function of $\tau$ strongly depends on the
expansion pattern, which, for early times, is dominated by longitudinal flow. Thus, in the initially
more compressed an later re-expanding best fit scenario the density is always higher than in the
Bjorken case, leading to a higher opacity of the medium. 
The effect of transverse flow is due to the influence of  Eq.\ref{E-Urs} --- and it can be made most
pronounced by increasing flow close to the surface. Using the formalism of \cite{Urs1,Urs2},
the center of the fireball is quite opaque and most measured jets originate close to the surface. Consequently, 
any effect proportional to the magnitude of $v_T$ is most pronounced if $v_T$ is enhanced
close to the surface. Therefore a flow profile $v_T \sim r^2$ shows more quenching than one $v_T \sim r$
\cite{Jet_Flow}.

In summary, we believe that given the large effects on the quenching power of the medium induced by flow,
it is not meaningful to present estimates of the gluon density in the initial state based on $R_{AA}$ alone
without taking the evolution of matter and the development of flow into account.

\section*{Acknowledgments}
I would like to thank J.~Ruppert, S.~A.~Bass and B.~M\"{u}ller for helpful discussions, comments and their
support during the preparation of this paper.

This work was supported by the DOE grant DE-FG02-96ER40945 and a Feodor
Lynen Fellowship of the Alexander von Humboldt Foundation.

\vfill\eject

\begin{thebibliography}{99}  
 \bibitem{Jet1}
  M.~Gyulassy and X.~N.~Wang,
  Nucl.\ Phys.\ B {\bf 420}, 583 (1994).


\bibitem{Jet2}
  R.~Baier, Y.~L.~Dokshitzer, A.~H.~Mueller, S.~Peigne and D.~Schiff,
  Nucl.\ Phys.\ B {\bf 484}, 265 (1997).
 

\bibitem{Jet3}
  B.~G.~Zakharov,
  JETP Lett.\  {\bf 65}, 615 (1997).

\bibitem{Jet4}
  U.~A.~Wiedemann,
  Nucl.\ Phys.\ B {\bf 588}, 303 (2000).


\bibitem{Jet5}
  M.~Gyulassy, P.~Levai and I.~Vitev,
  Nucl.\ Phys.\ B {\bf 594}, 371 (2001).


\bibitem{Jet6}
  X.~N.~Wang and X.~F.~Guo,
  Nucl.\ Phys.\ A {\bf 696}, 788 (2001).

\bibitem{Urs1}
  N.~Armesto, C.~A.~Salgado and U.~A.~Wiedemann,
  Phys.\ Rev.\ Lett.\  {\bf 93} (2004) 242301.

\bibitem{Urs2}
N.~Armesto, C.~A.~Salgado and U.~A.~Wiedemann,
hep-ph/0411341.

\bibitem{Mach}
  T.~Renk and J.~Ruppert,
  hep-ph/0509036.

\bibitem{Jet_Flow}
T.~Renk and J.~Ruppert,
hep-ph/0507075.

\bibitem{RenkSpectraHBT}
  T.~Renk,
  Phys.\ Rev.\ C {\bf 70} (2004) 021903.

\bibitem{STAR_RAA}
  J.~Adams {\it et al.}  [STAR Collaboration],
  Phys.\ Rev.\ Lett.\  {\bf 91} (2003) 172302.

\bibitem{JetScaling}
  C.~A.~Salgado and U.~A.~Wiedemann,
  Phys.\ Rev.\ Lett.\  {\bf 89}, 092303 (2002).

\bibitem{QuenchingWeights}
  C.~A.~Salgado and U.~A.~Wiedemann,
  Phys.\ Rev.\ D {\bf 68}, 014008 (2003).


\bibitem{Kari1}
  K.~J.~Eskola, H.~Honkanen, C.~A.~Salgado and U.~A.~Wiedemann,
  Nucl.\ Phys.\ A {\bf 747}, 511 (2005).

\bibitem{Kari2}
 K.~J.~Eskola and H.~Honkanen,
 Nucl.\ Phys.\ A {\bf 713}, 167 (2003).

\bibitem{CTEQ1}
  J.~Pumplin, D.~R.~Stump, J.~Huston, H.~L.~Lai, P.~Nadolsky and W.~K.~Tung,
  JHEP {\bf 0207}, 012 (2002).

\bibitem{CTEQ2}
  D.~Stump, J.~Huston, J.~Pumplin, W.~K.~Tung, H.~L.~Lai, S.~Kuhlmann and J.~F.~Owens,
  JHEP {\bf 0310}, 046 (2003).

\bibitem{NPDF}
  M.~Hirai, S.~Kumano and T.~H.~Nagai,
  Phys.\ Rev.\ C {\bf 70}, 044905 (2004).

\bibitem{KKP}
  B.~A.~Kniehl, G.~Kramer and B.~Potter,
  Nucl.\ Phys.\ B {\bf 582}, 514 (2000).


\end{thebibliography}
\end{document}